\documentclass[final,3p,times,twocolumn]{elsarticle}
\usepackage{amssymb}
\usepackage{float}
\usepackage[hyphens]{url}
\usepackage{algorithm}
\usepackage{algpseudocode}
\usepackage{multirow}
\usepackage{graphicx}

\begin{document}

\begin{frontmatter}

\title{Indriya: Building a Secure and Transparent Organ Donation System with Hyperledger Fabric}

\author[inst1]{Satyajit Ghosh}

\affiliation[inst1]{organization={Department of Computer Science and Engineering, Adamas University},%Department and Organization
            addressline={Barasat}, 
            city={Kolkata},
            postcode={700126}, 
            state={West Bengal},
            country={India}}
\ead{satyajit.ghosh.edu@gmail.com}
\author[inst1]{Mousumi Dutta}
\ead{mousumid721@gmail.com}
% \author[inst1]{Ratnadeep Dey}
\begin{abstract}
Recent technological advancements have led to the development of new methods for managing organ donation systems, which aim to overcome the limitations of traditional centralized systems. To achieve increased transparency, security, and efficiency in the organ donation process, blockchain technology is being proposed as a replacement for these centralized systems. However, most previous works on organ donation systems have focused on using Ethereum-based blockchain solutions, which offer limited control, a fixed set of consensus protocols, and no support for concurrent executions. In contrast, our work has utilized the Hyperledger Fabric framework to develop a network model of the organ donation system. We have designed and deployed a prototype system with smart contracts using Amazon Managed Blockchain Service. Additionally, we have built a client application that uses the Fabric SDK to interact with the network and perform various actions. To evaluate the performance of our system, we conducted extensive testing using the Hyperledger Caliper benchmarking tool. In our test bench, the system achieved a peak actual send rate of 389.1 transactions per second (TPS) for creating new records and 508.4 TPS for reading records. At a send rate of 800 TPS, the system took an average of 12.16 seconds to serve a request for creating a record and an average of 3.71 seconds to serve a request for reading a record. Future work is required to extend the functionalities of the system and identify potential endorsers and managers for this type of controlled blockchain network.
\end{abstract}


\end{frontmatter}

\section{Introduction}
Organ donation is a medical process that involves the transplantation of organs from one individual, the donor, to another individual, the recipient, who is in need of an organ. The organs can be donated by either living or deceased individuals. Living donors can donate organs such as a kidney, liver, or part of the lung, while deceased donors can donate multiple organs, including the heart, lungs, liver, pancreas, and kidneys. Organ donation is a critical procedure that can save or improve the quality of life for individuals suffering from a range of life-threatening or debilitating conditions.

Organ donation has a long and fascinating history that dates back to around 5000 years ago during the period of Rigveda, Ashwini Kumaras conducted both homo and hetero-transplantation procedures in ancient times \cite{singh_2017}. However, the practice of organ donation remained uncommon and largely experimental until the 20th century. In the 21st century, organ donation and transplantation have become more common and successful than ever before. However, the shortage of donor organs remains a significant challenge, leading to long waiting lists and high demand for innovative solutions to increase the availability of donor organs. According to World Health Organization(WHO) report more than 150000 solid organ transplants are performed worldwide annually \cite{director-general_2022}. 

The organ donation and transplantation process involves several steps, major ones are including the following:
\begin{itemize}
\item Determination of brain death: In order for a person to be considered for organ donation, they must first be declared brain dead. This step is not required for every donation although.
\item Evaluation of medical suitability: Once brain death has been declared, the person's medical history and current health status will be evaluated to determine whether they are a suitable organ donor. 
\item Matching with recipients: If the person is deemed a suitable organ donor, their organs will be matched with recipients who are in need of a transplant. This process involves matching the donor's blood and tissue type with the recipient's to ensure a successful transplant.
\item Organ retrieval \& transplant: Once a match has been found, the organs will be surgically removed from the donor's body and transplanted into the recipient's body.
\end{itemize}

Organ donation is a complex and sensitive process that involves numerous steps, from donor registration to organ transplantation. Managing the process requires careful coordination and communication among healthcare professionals, potential donors, and recipients. Software systems can be invaluable tools for managing the organ donation process. Software systems can help improve the efficiency of the process by automating many of the steps involved, reducing the likelihood of errors or delays. The organ donation process generates a significant amount of data, and a software system can help manage this data securely and make it easily accessible to authorized parties. This can improve the accuracy and completeness of the data, making it easier to make informed decisions about organ allocation. It can help track and report on each step of the process, ensuring that all regulatory requirements are met and that the process is transparent and accountable. Finally, a software system can help ensure that organs are allocated fairly and equitably to those who need them the most. By using objective criteria, such as medical urgency, rather than subjective factors such as race or socioeconomic status.

Blockchain-based organ donation systems offer several advantages over traditional database-based systems. One of the biggest advantages of blockchain-based systems is their enhanced security. Blockchain technology uses cryptography to ensure that data entered into the system is tamper-proof and cannot be altered or deleted. Decentralization makes it more resilient to system failures and reduces the risk of data loss or corruption. However, earlier implementations of blockchain-based organ donation systems have faced some significant limitations. Earlier blockchain-based organ donation systems were built on the Ethereum platform, which is known for its sequential transaction processing and lack of modularity. This made it difficult to scale the system to meet the demands of a large number of users, resulting in slow transaction times and reduced efficiency. In addition, earlier implementations of blockchain-based organ donation systems did not focus on developing a user-friendly front-end for the system and deployment. 

Our main contributions to this paper are:
\begin{itemize}
    \item We proposed the private and permissioned blockchain framework Hyperledger Fabric for the organ donation system.
    \item We have discussed the shortcomings of previously proposed currency-based frameworks i.e. Ethereum.
    \item We have developed smart contract algorithms and prototypes for an organ donation system using the Hyperledger Fabric framework and deployed it in AWS.
    \item We have compared Ethereum and Hyperledger and tested performance using the benchmarking tool, Hyperledger Caliper.
\end{itemize}
The rest of the paper is organized as follows:
Section II discusses the Existing Systems. Section III describes our Proposed Work. Section IV explains the Implementation. Section V shares results and discussion. Finally, Section VI concludes this paper.

\section{Existing Systems}
Organ donation is a crucial aspect of healthcare that saves lives and improves the recipients' quality of life. Technological advancements have developed new methods to manage organ donation systems more efficiently. In recent years, blockchain and non-blockchain-based organ donation systems have been explored and implemented to enhance the process of organ donation. Blockchain-based solutions for organ donation systems are mainly based on Ethereum and Hyperledger Fabric frameworks. The Ethereum-based solutions can be further divided into two types: public and private. Public Ethereum-based solutions use the public Ethereum network, which is accessible to anyone with an internet connection. Private Ethereum-based solutions, on the other hand, use a private Ethereum network that is only accessible to authorized participants. Hyperledger Fabric is a blockchain framework designed for enterprise use cases and is known for its flexibility, scalability, and security. Hyperledger Fabric-based solutions for organ donation systems aim to provide a secure and efficient platform for managing the entire organ donation process, from registration to transplantation.
\subsection{Non-Blockchain-based Solutions}
Non-blockchain-based organ donation systems, rely on traditional databases and techniques to manage the organ donation process. These systems typically use centralized databases and web-based platforms to manage the registration and distribution of organs.

A non-blockchain-based organ donation system is the United Network for Organ Sharing (UNOS), which is a non-profit organization responsible for managing the organ transplantation process in the United States. UNOS operates a web-based platform that connects organ donors and recipients and manages the allocation of organs based on medical criteria and geographical location \cite{unos_2023}.

Another non-blockchain-based organ donation system is the National Organ and Tissue Transplant Organization (NOTTO) in India, which operates a centralized database of organ donors and recipients, and manages the allocation of organs based on medical criteria \cite{notto}.

The Organ Procurement and Transplantation Network (OPTN) is a collaborative effort between public and private entities that brings together all parties involved. With the goal of addressing the shortage of organs for transplantation and improving the matching and placement process, the National Organ Transplant Act (NOTA) was passed by the U.S. Congress in 1984. The act established the OPTN as the national registry for organ matching \cite{NOTA}.

\subsection{Blockchain-based Solutions}
Traditional organ donation systems based on centralized databases have several disadvantages that hinder the efficiency and transparency of the organ donation process. One of the main challenges is a lack of transparency, which makes it difficult for stakeholders to access and verify the information. This can lead to questions about the fairness and ethical considerations of organ allocation. Furthermore, centralized databases are vulnerable to hacking, data breaches, and other security threats, which can compromise sensitive information about patients and donors. In addition, traditional systems often rely on manual processes for organ matching and placement, leading to inefficiencies and delays in the process. This, combined with inconsistent data quality and limited accessibility, can result in slow processing times and long wait times for patients in need of organ transplants. To address these limitations, it may be beneficial to adopt modern technology, such as blockchain-based systems, which offer increased transparency, security, and efficiency in the organ donation process.
\subsubsection{Ethereum-based Solutions}
Hawashin et. al. proposed a solution built on the private Ethereum blockchain and aims to provide a decentralized, secure, and trustworthy environment for the management of organ donations and transplants. The solution utilizes smart contracts to ensure data provenance by automatically recording events. The authors present six algorithms and provide details on their implementation, testing, and validation. The security of the proposed solution has been analyzed to ensure that smart contracts are protected against common attacks and vulnerabilities. A comparison has been made with other blockchain-based solutions that are currently available. The authors discuss the customization potential of the solution and its future improvement opportunities, including the development of an end-to-end DApp and deployment and testing on a real private Ethereum network. Additionally, it has been mentioned that the Quorum platform may provide better confidentiality as compared to the solution proposed in this study \cite{9787401}.

Ranjan et.al. proposed a decentralized and secure organ and tissue transplant web application. The aim of the DApp is to eliminate the need for third-party involvement in the organ transplantation process and to offer a cost-effective solution for patients. The authors implement the use of the IPFS (a distributed file server) to hash the details and Electronic Medical Records (EMR), resulting in reduced upload costs, as demonstrated in the results section of the paper. The proposed DApp aims to provide a transparent and secure platform for the management of organ and tissue transplants \cite{9066225}. The use of Ethereum coins in every transaction incurs a cost. Additionally, IPFS has limitations in terms of handling large files or high traffic volumes, which could impact its effectiveness for applications requiring quick and dependable access to vast amounts of data. The decentralized and distributed nature of IPFS, operating as a public P2P network, raises potential legal and regulatory concerns for the storage of confidential patient information \cite{8258226, https://doi.org/10.48550/arxiv.2202.06315}.

Wijayathilaka et.al. proposed a secure and smart blood and organ donation system that enables both patients and healthcare providers to access information about blood and organ processing records. The database is managed using blockchain technology, accessible only by authorized users. The system employs Ethereum Smart Contract to create a smart identity for all registered donors and uses a linear regression model to predict blood demand over the next ten years with a high accuracy of 0.998. This helps reduce blood shortages and waste. The system also leverages the global positioning system and K-Nearest Neighbors machine learning algorithm to find the best match between donors and recipients based on proximity. To assess donor awareness and attitudes towards organ donation, the system automatically sends questionnaires to registered users. The overall goal of this study is to provide a secure and transparent web application for blood and organ donation, streamlining the process in the Sri Lankan healthcare sector \cite{9357211}.

Apart from that Chaudhary et.al.\cite{9865787} also proposed an Ethereum-based organ donation system in their work.
\subsubsection{Hyperledger Fabric-based Solutions}

Lamba et. al. proposed a Hyperledger Fabric-based solution for organ donation systems. In this platform, the general public is able to view transactions carried out on the patient waiting list and organ donations without having access to the personal information of the participating individuals. This enhances the security and transparency of the system and helps to address the issue of organ black marketing and reducing costs in the organ donation transplantation process. The platform utilizes Fabric's immutable ledger to record all transactions and store replicated copies on multiple peers within the network, which can be the hospitals that oversee the transactions. The procedures for renewing the ledger require a quorum and a set number of validations to determine the validity of the transactions, making it a secure and transparent platform for the entire organ donation transplantation process\cite{8974526}.

Hai et. al. proposed a Hyperledger Fabric-based blood donation system. The authors present the design, development, and evaluation of the BloodChain system, a private blockchain-based blood donation network. The goal of the system is to secure the visibility of blood information and ensure the quality of blood for both donors and receivers. By recording and sharing information in a distributed ledger, BloodChain facilitates the efficient management of blood donations and classification. The information of all participants is transparent to one another in the same transaction, allowing for easy tracking of blood quality through metadata. The system also addresses the issue of information tampering and missing entries in traditional medical institutions, making the blood management process more transparent. Furthermore, BloodChain supports blood transactions among medical institutions and enables easy management of blood quality, benefiting both donors and receivers \cite{network2010002}.

Previous works on organ donation systems primarily focused on using the Ethereum blockchain solution and only a few utilized the Hyperledger Fabric framework. Despite the fact that Hyperledger Fabric offers several advantages for critical infrastructure systems. The majority of prior works concluded by merely constructing smart contracts and did not provide a complete implementation or deployment of the system. Our work addresses these shortcomings by providing a more comprehensive solution.
\section{Proposed Work}
\subsection{Background}
Blockchain is a decentralized, distributed ledger technology that allows for the secure and transparent recording of transactions. It uses a network of nodes to maintain a continuously growing list of ordered records, called blocks, which are linked and secured using cryptography. Another advantage of blockchain is that it is transparent and immutable. All transactions on the blockchain are recorded and visible to everyone on the network, which makes it easy to verify the legitimacy and provenance of the data. Additionally, once data has been written to the blockchain, it cannot be altered or deleted, which makes it a reliable and permanent record of transactions.
Public blockchains, also known as permissionless blockchains, are decentralized networks that are open to anyone. They allow anyone to join the network, view the transaction history, and participate in the consensus process to validate new transactions. Ethereum is an example of one such blockchain network. 

Private blockchains, also known as permissioned blockchains, are decentralized networks that are restricted to certain users. Unlike public blockchains, access to private blockchains is controlled by an authority, which determines who is allowed to join the network and participate in the consensus process. Private blockchains are often used by organizations that want to benefit from the security and transparency of blockchain technology without exposing their data to the public. Hyperledger Fabric is one such framework.
\subsection{Hyperledger Fabric Architecture}
The architecture of Hyperledger Fabric is modular and flexible, which allows for a high degree of customization and integration with existing systems. The components of the Hyperledger Fabric include Peer nodes that participate in the network and maintain a copy of the ledger. They can be divided into mainly three types: Endorser peers, Committing peers and Ordering peers. Endorser peers receive transactions from client applications, execute them, and return the results to the client. Committing peers receive the endorsed transactions, check them for validity, and then commit them to the ledger. Ordering peers package the endorsed transaction into blocks. Smart Contract or Chaincode is the business logic which is installed on the peers and executed by them when a client application submits a transaction. A membership services provider (MSP) manages the identities of the nodes on the network using Certificate Authorities (CA). It is used to authenticate and authorize the different nodes, ensuring that only authorized nodes can participate in the network and access its resources. The ledger of Hyperledger consists of two parts - a world state and a blockchain. World state uses a database to hold the current values of a set of ledger states so that queries can be performed efficiently. Hyperledger offers support for LevelDB and CouchDB. For complex queries, CouchDB is best suited\footnote{\url{https://hyperledger-fabric.readthedocs.io/en/latest/couchdb_as_state_database.html}}.

\subsection{Hyperledger Fabric Network Model}
%---------------------------------
\begin{figure*}
    \centering
  \includegraphics[height=10cm]{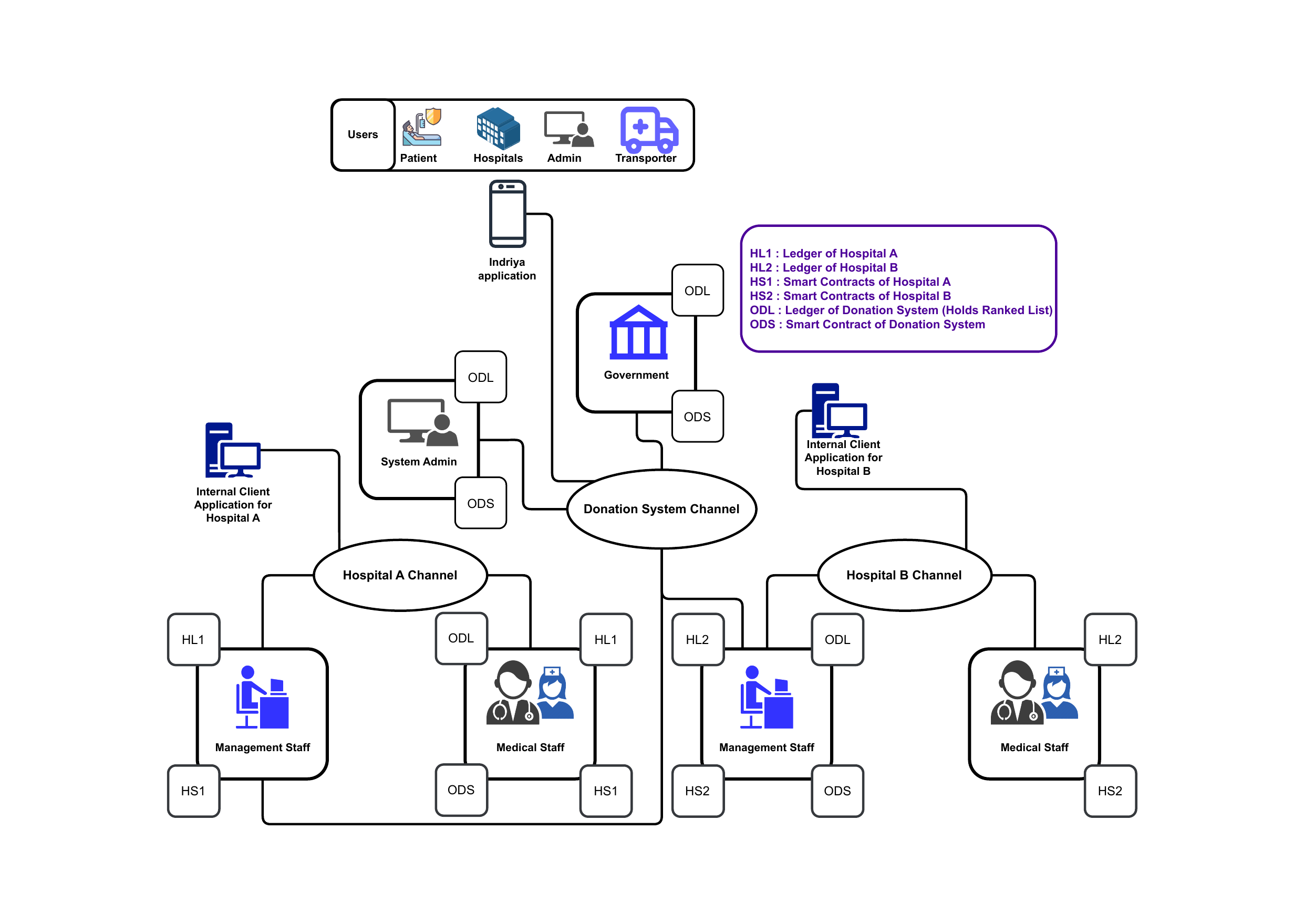}
  \caption{Proposed Network Model}
  \label{fig:model}
\end{figure*}
%--------------------------------

In our proposed system, the Hyperledger network is composed of peer nodes managed by different hospitals and government bodies. These hospitals can register patients and donors with basic medical details, and also have their own channels or subnetworks within the network. The medical, management staff and other departments of the hospital can have their own peers as per their organizational structure and rules. The hospital management can act as an endorser, ensuring that any tampering by the medical or any other department staff is prevented. In Hyperledger every peer holds the same local copy of the ledger and smart contract of the channel it has joined. If it has joined multiple channels then, it holds the ledger and smart contract of all the channels and can intercommunicate between the channel using smart contracts. The endorsing means that if a transaction is executed by any peer, then the endorser peer, will also execute that same transaction on its local copy of the ledger and if they both results in the same, then only the transaction is allowed to change the state of the network. That ensures a tempered ledger on any peer, doesn't impact the network. Likewise, in the main `Donation System channel' the Government can act as an endorser. The ledger of this channel can hold the records for donors or the ranked list. The internal ledger of the hospital can hold their patient data. The client applications are used to interact with the system. It can be extended to give access to the patients to see their own status of getting an organ. Government can also appoint an administrator and give him access through the client application to moderate the system. Figure \ref{fig:model} depicts the proposed model of the Hyperledger Fabric network.
\subsection{Actors}
%---------------------------------
\begin{figure*}
    \centering
  \includegraphics[height=10cm]{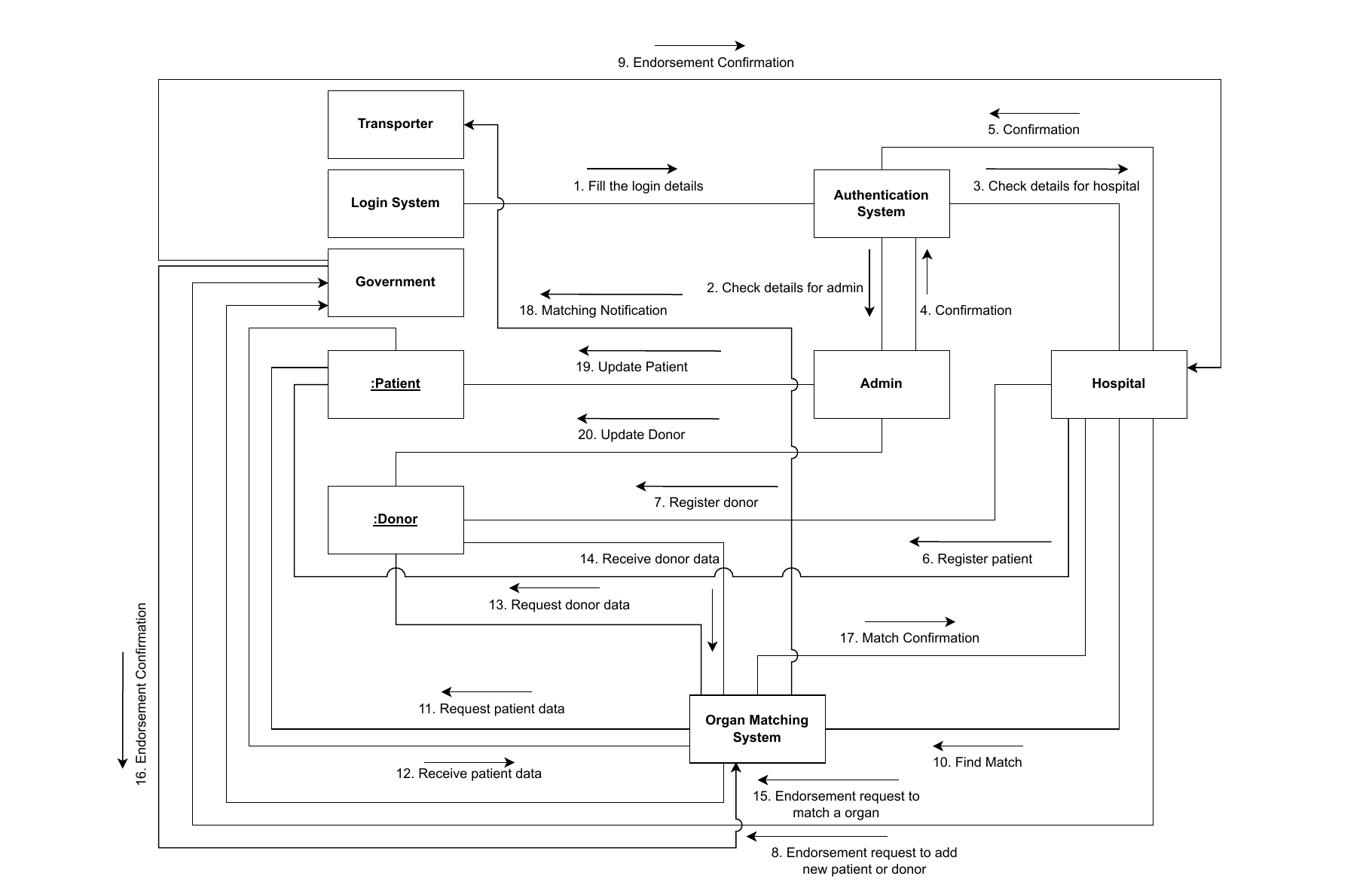}
  \caption{Collaboration diagram of Organ Donation System}
  \label{fig:collab}
\end{figure*}
%--------------------------------

The system has five main types of users: Patients, Hospitals, Administrators, Government and Transporters. Each of these user groups has specific roles and responsibilities in the system.
\subsubsection{Patients}
Patients can use the client application to access the system and view important information such as their current status in the organ transplant process, average waiting times, and whether an organ has been allocated to them or not.
\subsubsection{Hospitals}
Hospitals play a critical role in the system by adding patients or donors to the system and updating their information. They also use the match-making algorithm to find matching organs for their patients.
\subsubsection{Administrators}
Administrators are responsible for maintaining the legal requirements of the system and ensuring its security. They also monitor the system to make sure it is functioning correctly.
\subsubsection{Transporters}
Transporters play an important role in the successful transport of organs from one location to another. They are notified as soon as a match is found, and their job is to ensure that the organ reaches its intended recipient safely.
\subsubsection{Government}
The Government acts as an endorser of the system and serves as a monitoring body to ensure that everything is running smoothly.

By working together, these five user groups help to make the organ transplant process more efficient and effective. The collaboration between different actors of the system is depicted in Figure \ref{fig:collab}.
\section{Implemention}
Implementing a Hyperledger Fabric network involves several steps starting with setting up the network environment by installing the necessary software and tools such as Linux, Docker, Git and the Hyperledger Fabric platform. Next, the network components are defined by creating a network definition that specifies the participants, permissions, and relationships within the network. Then, a channel is created as a private and isolated communication path between network participants that allows for secure information sharing. The business logic of the network is governed by deploying chaincode, which is a smart contract that specifies the interactions between participants. After creating the network by deploying its components such as peers and orderers, thorough testing should be done to ensure it functions as expected. Finally, the network is deployed to the production environment and made available to network participants.
\subsection{Smart Contracts and Algorithms}

\begin{table*}[]
\caption{Smart contract methods}
\label{tab:smart_contract}
\resizebox{\textwidth}{!}{%
\begin{tabular}{|l|l|}
\hline
\textbf{Method Name}               & \textbf{Description}                                                         \\ \hline
addPatient(Patient Details)        & It is used to add a new patient in the system by hospitals                   \\ \hline
addDonor(Donor Details)            & It is used to add new donor in the system by hospitals                       \\ \hline
getAllPatients()                   & It is used to fetch all the registered patients in the system by admin       \\ \hline
getAllDonors()                     & It is used to fetch all the registered donors in the system by admin         \\ \hline
getPatient(Patient\_ID)            & It is used to fetch one particular patient details                           \\ \hline
getDonor(Donor\_ID)                & It is used to fetch one particular donor details                             \\ \hline
deletePatient(Patient\_ID)         & It is used to delete a patient                                               \\ \hline
deleteDonor(Donor\_ID)             & It is used to delete a donor                                                 \\ \hline
getMyPatients(Hospital\_ID)        & It is used by a hospital to fetch their all patients                         \\ \hline
getMyDonors(Hospital\_ID)          & It is used by a hospital to fetch their all donors                           \\ \hline
findMatch(Patient\_ID)             & It is used to search for matching donors for a patient by patient's hospital \\ \hline
selectMatch(Patient\_ID,Donor\_ID) & It is used to finalize matching donor with the patient                       \\ \hline
\end{tabular}%
}
\end{table*}

We have prepared smart contracts for the admin and hospitals to execute in the system to perform various tasks as described in Table \ref{tab:smart_contract}.

%--------------Algorithms---------------%
\begin{algorithm}
\caption{Adding new Patient or Donor}
\label{alg:add_new_patient}
\begin{algorithmic}[1]
\Require [$PatientID$/$DonorID$], $firstName$, $lastName$, $age$, $phoneNumber$, $address$, $organRequired$, $bloodgroup$, $gender$, $medhistory$
\Ensure $ID$ not exists
\State Create record $r$ of type Patient or Donor
\State Set $r.ID=ID$
\State Set $r.firstName=firstName$
\State Set $r.lastName=lastName$
\State Set $r.age=age$
\State Set $r.phoneNumber=phoneNumber$
\State Set $r.address=address$
\State Set $r.organRequired=organRequired$
\State Set $r.bloodgroup=bloodgroup$
\State Set $r.gender=gender$
\State Set $r.medhistory=medhistory$
\State Submit $r$ to add to the Blockchain.
\end{algorithmic}
\end{algorithm}
%------------------Match Making Algorithm------------
\begin{algorithm}
\caption{Patient-Donor Matchmaking}
\label{alg:pat_donor_match}
\begin{algorithmic}[1]
\Require $organRequired$, $bloodgroup$, $gender$
\State Let $D\textsubscript{all}=getAll(Donor)$
\State Let $matches=[]$
  \For{\texttt{i} in $D\textsubscript{all}$}
    \texttt{\If{$i.bloodgroup$==$bloodgroup$ and $i.organRequired$==$organRequired$ and $i.gender$==$gender$ }
    \State $matches.push(i)$ .
    \EndIf}
    \EndFor
\State \textbf{Return} $matches$
\end{algorithmic}
\end{algorithm}

\begin{algorithm}
\caption{Select a Matching Donor}
\label{alg:final_match}
\begin{algorithmic}[1]
\Require $DonorID$, $PatientID$, $matches$
\Ensure $DonorID$ is present in $matches$ for the $PatientID$
\begin{Statex}
/* $r\textsubscript{d}$ is a object of type DonorRecord and
$r\textsubscript{p}$ is a object of type PatientRecord */
\end{Statex}
\State $r\textsubscript{d}=get(DonorID)$ 
\State $r\textsubscript{p}=get(PatientID)$
\State $r\textsubscript{d}.match=$=$PatientID$
\State $r\textsubscript{p}.match=$=$DonorID$
\State Submit $r\textsubscript{d}$ and $r\textsubscript{p}$ to update the blockchain ledger.
\end{algorithmic}
\end{algorithm}

%---------------------------------
As earlier research efforts on blockchain-based organ donation systems, already focused on designing algorithms, our work emphasizes the implementation and deployment of a system using the Hyperledger Fabric platform. We have developed a core set of minimal algorithms that are optimized for use on this platform, which we believe will enhance the scalability, security, and efficiency of the organ donation process. 

Algorithm \ref{alg:add_new_patient} describes the \texttt{addPatient()} and \texttt{addDonor()} method. It is used to register a patient or a donor with their personal and medical details in the system.

 Next, Algorithm \ref{alg:pat_donor_match} describes the \texttt{findMatch()} method. It checks for matching donors in the list for a patient. 
 
 Algorithm \ref{alg:final_match} describes the \texttt{selectMatch()} method. It is used to cross-update the match information in their profile.

\subsection{Back-end}
Hyperledger backend can be written in several programming languages like Node.js, Java, Go etc. We have used Node.js along with Express.js for the back-end implementation as it is helpful for building fast and scalable applications in less time.
Node.js is a widely used JavaScript runtime environment that can act as the backend for a Hyperledger Fabric network. Interaction between the Node.js backend and the Hyperledger Fabric network is facilitated through the Hyperledger Fabric client SDK. The SDK offers a suite of APIs that permit the backend to connect with the network, carry out transactions, query the ledger, and listen to events.

To interact with the Hyperledger Fabric network, the Node.js backend must first establish a connection to one of the peers in the network. Next, an identity must be set up, which requires a certificate and a private key. With the connection established and the identity in place, the backend can use the SDK to initiate transactions on the network, such as creating, updating, or retrieving information about assets. Additionally, the backend can query the ledger for information and listen for events emitted by the network, such as the completion of a transaction. Express.js can be used to create RESTful APIs that interact with the Hyperledger Fabric network. It provides a range of advanced features, including support for middleware, templating engines, and routing.
\subsection{Front-end}
To build the front-end of the application we have used Embedded JavaScript(EJS). It allows you to generate dynamic HTML pages by embedding JavaScript code in the template. When the template is rendered by the server, JavaScript is executed and replaced with dynamic content. EJS is particularly useful for web development, where it's often necessary to generate HTML pages on the server side and send them to the client.
\subsection{Deployment}
Hyperledger Fabric is a flexible and scalable blockchain platform that can be deployed in various ways to meet the needs of different organizations and use cases. Local deployment of Hyperledger Fabric can be done on a single machine for testing and development purposes and is suitable for smaller organizations or for evaluating the platform. On the other hand, multi-node deployment of Hyperledger Fabric involves multiple nodes across multiple machines and is suitable for larger organizations with complex use cases and higher security requirements. Cloud deployment of Hyperledger Fabric on platforms such as Amazon Web Services (AWS), Microsoft Azure, or Google Cloud Platform provides a scalable and flexible infrastructure that can be easily managed and maintained. A hybrid deployment, combining local and cloud deployment methods, is suitable for organizations with varying needs, such as some components being deployed locally for testing and development while others are deployed on the cloud for production. Hyperledger Fabric can also be deployed as a managed service offered by blockchain providers, freeing up the organization's resources to focus on the development of its applications.
\subsubsection{Amazon Managed Blockchain Service}
Amazon Managed Blockchain is a fully managed service offered by Amazon Web Services (AWS) that makes it easy to create and manage scalable blockchain networks. The service provides a scalable infrastructure to host Hyperledger Fabric networks, enabling organizations to focus on building their applications while leaving the management of the underlying blockchain network to AWS.

Amazon Managed Blockchain offers a secure and reliable environment for blockchain networks, as it provides automatic backups and disaster recovery options, which reduce the risk of data loss or downtime. The service also provides network management and monitoring tools, making it easier to manage and monitor network performance.

One of the key benefits of Amazon Managed Blockchain is its scalability, which allows organizations to easily add or remove nodes as needed, making it possible to manage the network size and capacity based on changing demands. The service also provides support for multiple members and enables users to invite other AWS accounts to join the network, making it possible to build and manage large-scale, multi-party blockchain networks.

We have this managed blockchain service, for deploying our organ donation system. At first, we created a private network of Hyperledger Fabric 2.2 with an approval threshold greater than 50\% and a proposal duration of 24 hours and added a member. Next, created a VPC endpoint for the member. This allows the Amazon EC2 instance that we use as a Hyperledger Fabric client to interact with the Hyperledger Fabric endpoints that Amazon Managed Blockchain exposes for our members. After that, we created two peer nodes in the network. Next, we verified the Fabric CA and VPC endpoint connection. 

After that, we created a channel. In the context of a channel, a ledger is present. If all members operate within a common channel, the ledger can be shared across the network. Alternatively, the privacy of a channel can be maintained by limiting its participants to a specific set of members. These members can either belong to the same AWS account or be invited from other AWS accounts. We described our channel configuration in a \texttt{configtx.yaml} file and joined our peer nodes with the channel.
%---------------------------------
\begin{figure*}
    \centering
  \includegraphics[height=10cm]{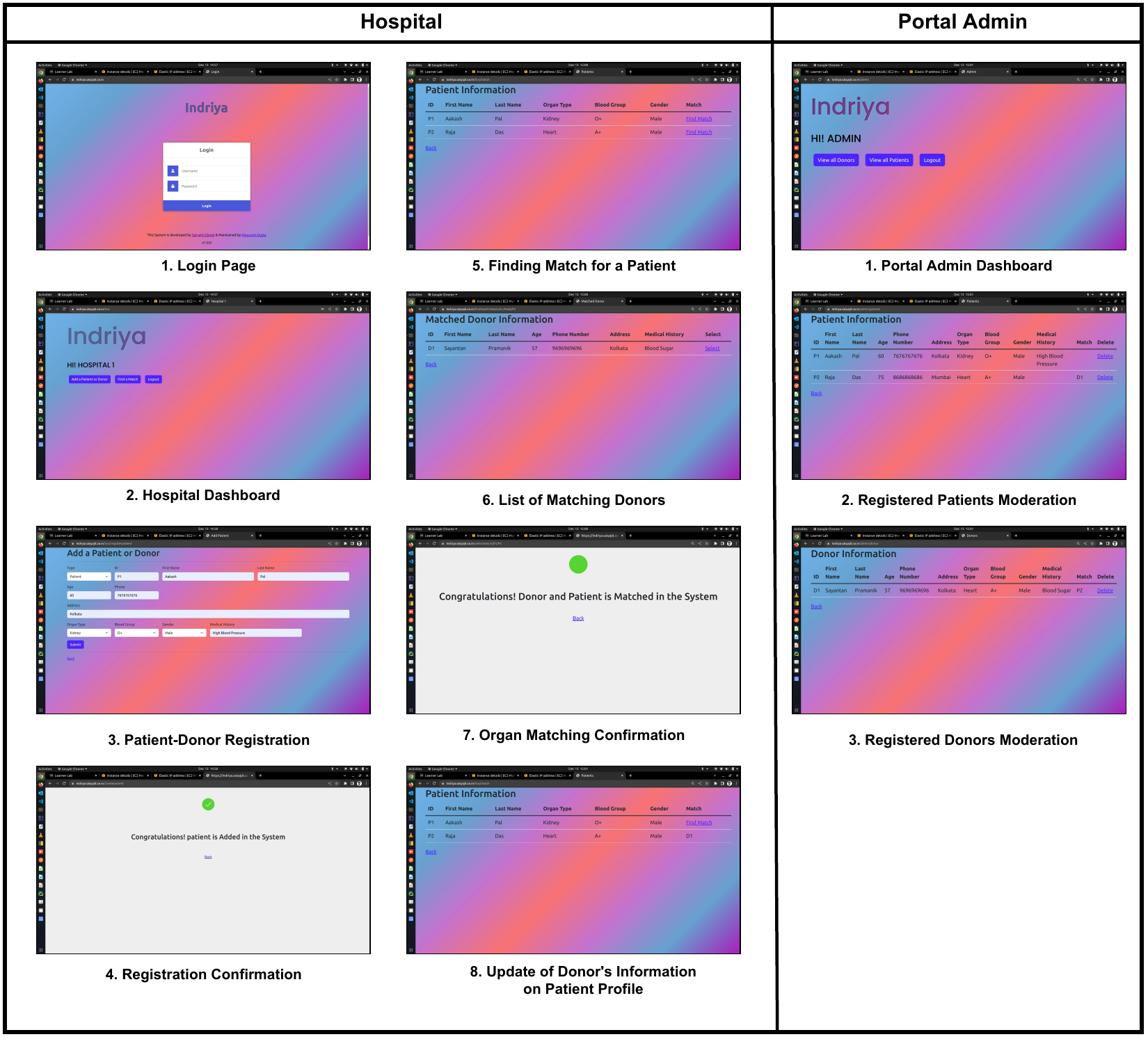}
  \caption{Client Application running on AWS}
  \label{fig:client}
\end{figure*}
%--------------------------------
%---------------------------------
\begin{figure*}
    \centering
  \includegraphics[height=10cm]{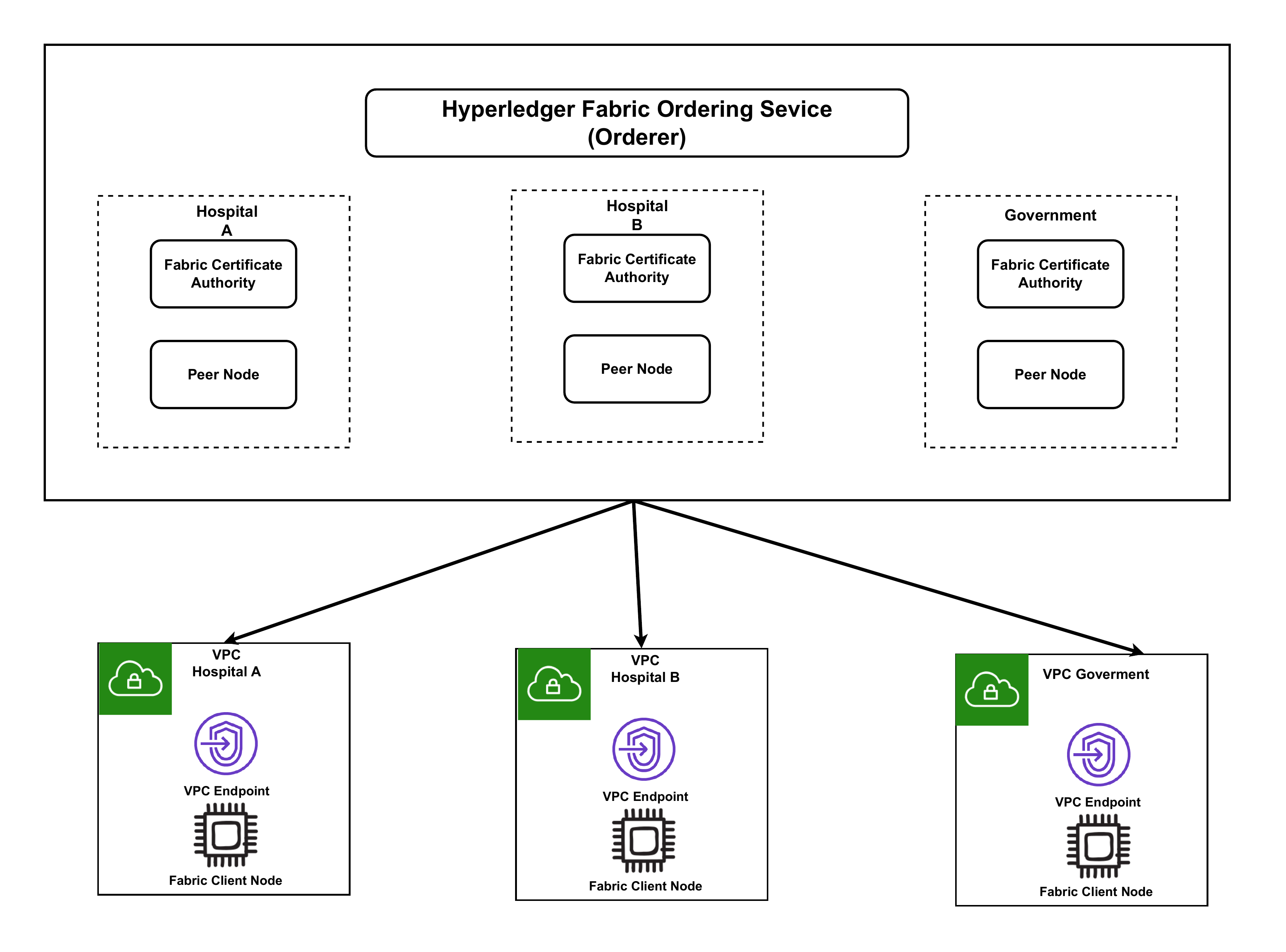}
  \caption{Hyperledger Network on Amazon Managed Blockchain Service}
  \label{fig:blackbox}
\end{figure*}
%--------------------------------

Next, we installed our chaincode in the channel, after packaging and verifying it.

Next, we installed the frontend dependencies on the client VPC endpoint and started our ExpressJS server. To ensure the secure transmission between the user and server, we generated SSL certificates on the client node, with the help of \texttt{Certbot}\footnote{\url{https://certbot.eff.org/}}. Figure \ref{fig:client} displays our client application and The black box view of the deployed system is depicted in Figure \ref{fig:blackbox}.

\section{Results and Discussion}
\subsection{Comparison with Ethereum}
Ethereum blockchain in comparison to Hyperledger Fabric for private systems is that Ethereum is a public blockchain and it may not be suitable for handling confidential and sensitive information related to organ donation and transplantation. This is because Ethereum is a permissionless blockchain, which means that any participant can access and view the transactions in the blockchain. On the other hand, even if we use the private chain in Ethereum, it may not provide the benefits of modular network design and concurrent execution, as well as fine-tuned access control for different system users, compared to the Hyperledger Fabric framework for private systems. 

Earlier studies found that Ethereum with the Proof-of-Work consensus model has higher latency and lower throughput than the Hyperledger Fabric 0.6 version \cite{8038517} but when we attempted to compare the performance of Ethereum and Hyperledger Fabric 2.2 by implementing the same set of smart contracts, however, we discovered that this comparison would not be useful or appropriate  as key factors such as batch size, batch time, and consensus mechanisms differ between the two platforms. In Ethereum, the gas limit determines the key parameter, whereas, in Hyperledger Fabric, it can be configured using data size or the number of transactions per batch. Ethereum uses a proof-of-work or proof-of-stake consensus mechanism, whereas Hyperledger Fabric uses a permissioned blockchain architecture and can use various consensus mechanisms such as PBFT or RAFT. Additionally, Ethereum has a global state and all transactions are processed in the same order, whereas Hyperledger Fabric has a modular architecture with different channels, each with its own private state, and transactions can be processed concurrently.
\subsection{Comparison with Existing Systems}

\begin{table*}[]
\caption{Comparison of existing solutions with our solution}
\label{tab:compare}
\resizebox{\textwidth}{!}{%
\begin{tabular}{|l|l|l|l|l|l|}
\hline
\textbf{Features}       & \textbf{\begin{tabular}[c]{@{}l@{}}UNOS\cite{unos_2023}\\ NOTTO\cite{notto}\\ OPTN\cite{NOTA}\end{tabular}} & \textbf{Hawashin et. al.\cite{9787401}} & \textbf{\begin{tabular}[c]{@{}l@{}}Ranjan et.al.\cite{9066225}\\ Chaudhary et.al.\cite{9865787}\end{tabular}} & \textbf{\begin{tabular}[c]{@{}l@{}}Lamba et. al.\cite{8974526}\\ Hai et. al.\cite{network2010002}\end{tabular}} & \textbf{Our Solution}   \\ \hline
Platform                & Database                                                             & Ethereum                  & Ethereum                                                                          & Hyperledger Fabric                                                           & Hyperledger Fabric      \\ \hline
Mode of Operation       & Private                                                              & Private                   & Public                                                                            & Private                                                                      & Private                 \\ \hline
Consensus mechanism     & Nil                                                                  & Proof-of-Work             & Proof-of-Work                                                                     & Raft                                                                         & Raft                    \\ \hline
Governance and control  & Yes                                                                  & Limited Control           & Limited Control                                                                   & Yes                                                                          & Yes                     \\ \hline
Single point of failure & Yes                                                                  & No                        & No                                                                                & No                                                                           & No                      \\ \hline
Transparency            & No                                                                   & Yes                       & Yes                                                                               & Yes                                                                          & Yes                     \\ \hline
Implementation          & Yes                                                                  & Only Smart Contracts      & Only Smart Contracts                                                              & Only Smart Contracts                                                         & Smart Contracts and GUI \\ \hline
Deployment              & Yes                                                                  & No                        & No                                                                                & No                                                                           & Yes                     \\ \hline
\end{tabular}%
}
\end{table*}

Table \ref{tab:compare} provides a comparative analysis of our proposed solution with existing solutions, based on key parameters such as the blockchain platform used, deployment, and mode of operation. Our solution leverages the Hyperledger Fabric platform, which has been used previously by \cite{8974526} and \cite{network2010002}. However, our solution distinguishes itself by incorporating a frontend and addressing deployment considerations, which were not covered in previous work. On the other hand, some existing implementations rely on the Ethereum blockchain framework, which can provide limited control, a fixed set of consensus protocols, and no support for concurrent executions. Overall, our proposed solution offers a more comprehensive and advanced approach to the problem at hand, leveraging the latest technologies and addressing key limitations of previous work.

\subsection{Performance Evaluation}
We tested the performance of our prototype network using Hyperledger Caliper.
Hyperledger Caliper is a performance benchmarking tool for Hyperledger blockchain platforms. It is used to measure the performance and scalability of different Hyperledger platforms, including Hyperledger Fabric, Hyperledger Sawtooth, and Hyperledger Iroha. Caliper provides a flexible and extensible architecture that allows developers to write their own custom test cases, which can be tailored to meet their specific performance requirements. Additionally, Caliper provides a set of predefined test cases that can be used to measure the performance of various blockchain functionalities. 

We conducted testing on a system equipped with an AMD Ryzen 3 5300U processor with a base frequency of 2.6GHz and a maximum frequency of 3.8GHz, and 8GB of DDR4-3200 RAM. The blockchain network was configured with a RAFT ordering service and two organizations, each with two peers. Endorsements could be performed by any peer within the organization.

The Hyperledger Caliper tool measures a blockchain's performance by recording statistics related to the processing of transactions. The following parameters are commonly recorded:
\begin{itemize}
    \item 
    Throughput: This metric measures the average number of transactions that can be processed per second.
    \item 
    Latency: This metric measures the time it takes from when a transaction is issued until the response is received.
    \item 
    Successful/Fail: This metric records the number of successful and failed transactions during the measurement period.
    \item 
    Resources: This metric provides information about the resources utilized during the measurement, such as memory usage and CPU utilization.
\end{itemize}

We have tested the network to create a new patient record and to read patient data.

\subsubsection{Experiment 1}
In the first experiment, described in Table \ref{tab:exp1}, we conducted a series of transactions using the Fixed-Load configuration on Hyperledger Caliper. The purpose of this configuration was to optimize throughput by setting a predefined set of backlog transactions, also referred to as Transaction Load. Specifically, we used transaction loads of 25, 50, 75, and 100 to perform 1000 transactions. Our results showed that the system achieved a peak throughput of 227.9 transactions per second (TPS) when creating new records, and 248.4 TPS when reading records, at a transaction load of 100. 

\subsubsection{Experiment 2}

As outlined in Table \ref{tab:exp2}, our second experiment involved the use of Hyperledger Caliper's Fixed-rate configuration, which involves sending transaction requests to the network at a fixed configured send rate. For this experiment, we configured Caliper to send requests at rates of 200, 400, 600, and 800 transactions per second (TPS). The results showed that the system achieved a peak actual send rate of 389.1 TPS for creating new records, and 508.4 TPS for reading records. At a send rate of 800 TPS, it took an average of 12.16 seconds to serve a request for creating a record, and an average of 3.71 seconds to serve a request for reading a record. 

These findings suggest that our system is capable of delivering fast and efficient performance when handling high transaction loads, making it a viable solution.

\begin{table*}[h]
\caption{Results from Experiment 1}
\label{tab:exp1}
\resizebox{\textwidth}{!}{%
\begin{tabular}{|l|l|l|l|l|l|l|}
\hline
Name                                                                     & Transaction Load & Send Rate (TPS) & Max Latency (s) & Min Latency (s) & Avg Latency (s) & Throughput (TPS) \\ \hline
\multirow{4}{*}{\begin{tabular}[c]{@{}l@{}}Create\\ Record\end{tabular}} & 25               & 46.1            & 1.87            & 0.06            & 0.27            & 46.0             \\ \cline{2-7} 
                                                                         & 50               & 77.7            & 1.86            & 0.06            & 0.32            & 76.5             \\ \cline{2-7} 
                                                                         & 75               & 75.7            & 1.87            & 0.06            & 0.50            & 75.4             \\ \cline{2-7} 
                                                                         & 100              & 90.2            & 1.88            & 0.07            & 0.59            & 86.2             \\ \hline
\multirow{4}{*}{Read Record}                                             & 25               & 228.3           & 0.09            & 0.01            & 0.04            & 227.9            \\ \cline{2-7} 
                                                                         & 50               & 231.2           & 0.17            & 0.01            & 0.08            & 230.7            \\ \cline{2-7} 
                                                                         & 75               & 247.9           & 0.21            & 0.01            & 0.11            & 247.4            \\ \cline{2-7} 
                                                                         & 100              & 248.9           & 0.28            & 0.01            & 0.14            & 248.4            \\ \hline
\end{tabular}%
}
\end{table*}

\begin{table*}[h]
\caption{Results from Experiment 2}
\label{tab:exp2}
\resizebox{\textwidth}{!}{%
\begin{tabular}{|l|l|l|l|l|l|l|}
\hline
Name                                                                     & Configured Send Rate(TPS) & Actual Send Rate (TPS) & Max Latency (s) & Min Latency (s) & Avg Latency (s) & Throughput (TPS) \\ \hline
\multirow{4}{*}{\begin{tabular}[c]{@{}l@{}}Create\\ Record\end{tabular}} & 200                       & 196.8                  & 12.77           & 0.49            & 6.32            & 117.8            \\ \cline{2-7} 
                                                                         & 400                       & 387.1                  & 17.98           & 6.18            & 12.13           & 105.8            \\ \cline{2-7} 
                                                                         & 600                       & 378.9                  & 18.27           & 5.79            & 12.15           & 104.9            \\ \cline{2-7} 
                                                                         & 800                       & 389.1                  & 18.19           & 6.00            & 12.16           & 106.5            \\ \hline
\multirow{4}{*}{Read Record}                                             & 200                       & 200.1                  & 0.02            & 0.01            & 0.01            & 200.0            \\ \cline{2-7} 
                                                                         & 400                       & 391.3                  & 4.82            & 0.18            & 2.31            & 376.5            \\ \cline{2-7} 
                                                                         & 600                       & 508.4                  & 6.36            & 0.02            & 4.27            & 303.3            \\ \cline{2-7} 
                                                                         & 800                       & 505.2                  & 6.00            & 2.13            & 3.71            & 324.9            \\ \hline
\end{tabular}%
}
\end{table*}

\section{Conclusion and Future Work}
The use of a permissioned blockchain, as opposed to a currency-based blockchain, offers several key benefits for our organ donation system. First, it allows us to control who has access to the network, ensuring that only authorized individuals are able to view and update the information on the blockchain. Second, it allows us to tailor the network to our specific needs. Our Hyperledger-based organ donation system offers a number of potential benefits, including improved transparency, and security. In this paper, We have discussed the implementation of our prototype system using Amazon Managed Blockchain for a secure and scalable blockchain network. We have also presented our underlying smart contract algorithms and compared our system with existing systems. As private blockchain networks are managed by authorities, it requires more future inspection to decide who will be the potential endorsers and managers for this type of system to improve and maintain the transparency and accountability of the system. Further work is needed to extend the front-end
application’s functionalities. All the supplementary files and source codes are available from \url{https://github.com/MousumiDutta2000/Indriya}.
\section*{Acknowledgment}
We would like to express my sincere gratitude to Professor Ratnadeep Dey, Dept. of Computer Science and Engineering, Adamas University for his invaluable guidance and supervision throughout this research project. His extensive knowledge, insightful comments, and constructive feedback were instrumental in shaping this paper and improving its quality.
\bibliographystyle{elsarticle-num} 
\bibliography{cas-refs}

\begin{thebibliography}{10}
\expandafter\ifx\csname url\endcsname\relax
  \def\url#1{\texttt{#1}}\fi
\expandafter\ifx\csname urlprefix\endcsname\relax\def\urlprefix{URL }\fi
\expandafter\ifx\csname href\endcsname\relax
  \def\href#1#2{#2} \def\path#1{#1}\fi

\bibitem{singh_2017}
V.~Singh, Sushruta: The father of surgery, National Journal of Maxillofacial
  Surgery 8~(1) (2017) 1–3.
\newblock \href {https://doi.org/10.4103/njms.njms\_33\_17}
  {\path{doi:10.4103/njms.njms\_33\_17}}.

\bibitem{director-general_2022}
Director-General,
  \href{https://apps.who.int/gb/ebwha/pdf_files/WHA75/A75_41-en.pdf}{Human
  organ and tissue transplantation - world health organization} (Apr 2022).
\newline\urlprefix\url{https://apps.who.int/gb/ebwha/pdf_files/WHA75/A75_41-en.pdf}

\bibitem{unos_2023}
\href{https://unos.org/}{United network for organ sharing: Us organ
  transplantation} (Feb 2023).
\newline\urlprefix\url{https://unos.org/}

\bibitem{notto}
\href{https://notto.gov.in/}{National organ \& tissue transplant organisation}.
\newline\urlprefix\url{https://notto.gov.in/}

\bibitem{NOTA}
\href{https://optn.transplant.hrsa.gov/about/history-nota/}{History and nota -
  optn}.
\newline\urlprefix\url{https://optn.transplant.hrsa.gov/about/history-nota/}

\bibitem{9787401}
D.~Hawashin, R.~Jayaraman, K.~Salah, I.~Yaqoob, M.~C.~E. Simsekler,
  S.~Ellahham, Blockchain-based management for organ donation and
  transplantation, IEEE Access 10 (2022) 59013--59025.
\newblock \href {https://doi.org/10.1109/ACCESS.2022.3180008}
  {\path{doi:10.1109/ACCESS.2022.3180008}}.

\bibitem{9066225}
P.~Ranjan, S.~Srivastava, V.~Gupta, S.~Tapaswi, N.~Kumar, Decentralised and
  distributed system for organ/tissue donation and transplantation, in: 2019
  IEEE Conference on Information and Communication Technology, 2019, pp. 1--6.
\newblock \href {https://doi.org/10.1109/CICT48419.2019.9066225}
  {\path{doi:10.1109/CICT48419.2019.9066225}}.

\bibitem{8258226}
Y.~Chen, H.~Li, K.~Li, J.~Zhang, An improved p2p file system scheme based on
  ipfs and blockchain, in: 2017 IEEE International Conference on Big Data (Big
  Data), 2017, pp. 2652--2657.
\newblock \href {https://doi.org/10.1109/BigData.2017.8258226}
  {\path{doi:10.1109/BigData.2017.8258226}}.

\bibitem{https://doi.org/10.48550/arxiv.2202.06315}
T.~V. Doan, Y.~Psaras, J.~Ott, V.~Bajpai,
  \href{https://arxiv.org/abs/2202.06315}{Towards decentralised cloud storage
  with ipfs: Opportunities, challenges, and future directions} (2022).
\newblock \href {https://doi.org/10.48550/ARXIV.2202.06315}
  {\path{doi:10.48550/ARXIV.2202.06315}}.
\newline\urlprefix\url{https://arxiv.org/abs/2202.06315}

\bibitem{9357211}
P.~Wijayathilaka, P.~P. Gamage, K.~De~Silva, A.~Athukorala,
  K.~Kahandawaarachchi, K.~Pulasinghe, Secured, intelligent blood and organ
  donation management system - “lifeshare”, in: 2020 2nd International
  Conference on Advancements in Computing (ICAC), Vol.~1, 2020, pp. 374--379.
\newblock \href {https://doi.org/10.1109/ICAC51239.2020.9357211}
  {\path{doi:10.1109/ICAC51239.2020.9357211}}.

\bibitem{9865787}
N.~Chaudhary, S.~S. Manvi, N.~Koul, Organ bank based on blockchain, in: 2022
  IEEE International Conference on Electronics, Computing and Communication
  Technologies (CONECCT), 2022, pp. 1--5.
\newblock \href {https://doi.org/10.1109/CONECCT55679.2022.9865787}
  {\path{doi:10.1109/CONECCT55679.2022.9865787}}.

\bibitem{8974526}
R.~Lamba, Y.~Gupta, S.~Kalra, M.~Sharma, Preventing waiting list manipulation
  and black marketing of donated organs through hyperledger fabric, in: 2019
  International Conference on Computing, Communication, and Intelligent Systems
  (ICCCIS), 2019, pp. 280--285.
\newblock \href {https://doi.org/10.1109/ICCCIS48478.2019.8974526}
  {\path{doi:10.1109/ICCCIS48478.2019.8974526}}.

\bibitem{network2010002}
H.~T. Le, T.~T.~L. Nguyen, T.~A. Nguyen, X.~S. Ha, N.~Duong-Trung,
  \href{https://www.mdpi.com/2673-8732/2/1/2}{Bloodchain: A blood donation
  network managed by blockchain technologies}, Network 2~(1) (2022) 21--35.
\newblock \href {https://doi.org/10.3390/network2010002}
  {\path{doi:10.3390/network2010002}}.
\newline\urlprefix\url{https://www.mdpi.com/2673-8732/2/1/2}

\bibitem{8038517}
S.~Pongnumkul, C.~Siripanpornchana, S.~Thajchayapong, Performance analysis of
  private blockchain platforms in varying workloads, in: 2017 26th
  International Conference on Computer Communication and Networks (ICCCN),
  2017, pp. 1--6.
\newblock \href {https://doi.org/10.1109/ICCCN.2017.8038517}
  {\path{doi:10.1109/ICCCN.2017.8038517}}.

\end{thebibliography}
\end{document}